\documentclass[%
 twocolumn, 12pt,
 amssymb,
 aps,
]{aastex631}

\usepackage{graphicx}
\usepackage{dcolumn}
\usepackage{bm}
\usepackage{floatrow}
\usepackage{hyperref}

\usepackage{graphicx}
\usepackage{natbib}
\usepackage{subcaption}
\usepackage[fleqn]{amsmath}
\begin{document}

\title{Virgo Filaments IV: Using WISE to Measure the Modification of Star-Forming Disks in the Extended Regions Around the Virgo Cluster}

\author[0009-0005-0303-0330]{Kim Conger}
\altaffiliation[Email: ]{kconger@ku.edu}
\affiliation{University of Kansas, Department of Physics and Astronomy, 1251 Wescoe Hall Drive, Room 1082, Lawrence, KS 66049, USA\\}

\author[0000-0001-5851-1856]{Gregory Rudnick}
\affiliation{University of Kansas, Department of Physics and Astronomy, 1251 Wescoe Hall Drive, Room 1082, Lawrence, KS 66049, USA\\}

\author[0000-0001-8518-4862]{Rose A. Finn}
\affiliation{Department of Physics and Astronomy, Siena College, 515 Loudon Road, Loudonville, NY 12211, USA\\}

\author[0000-0001-6831-0687]{Gianluca Castignani}
\affiliation{INAF - Osservatorio di Astrofisica e Scienza dello Spazio di Bologna, via Gobetti 93/3, I-40129, Bologna, Italy\\}

\author{John Moustakas}
\affiliation{Department of Physics and Astronomy, Siena College, 515 Loudon Road, Loudonville, NY 12211, USA\\}

\author[0000-0003-0980-1499]{Benedetta Vulcani}
\affiliation{INAF- Osservatorio astronomico di Padova, Vicolo Osservatorio 5, I-35122 Padova, Italy\\}

\author[0009-0001-1809-4821]{Daria Zakharova}
\affiliation{Dipartimento di Fisica e Astronomia Galileo Galilei, Universit\`a degli studi di Padova, Vicolo dell’Osservatorio, 3, I-35122 Padova, Italy
}
\affiliation{INAF- Osservatorio astronomico di Padova, Vicolo Osservatorio 5, I-35122 Padova, Italy\\}

\author[0000-0003-3864-068X]{Lizhi Xie}
\affiliation{Tianjin Astrophysics Center, Tianjin Normal University, Binshuixidao 393, 
300387 Tianjin, China\\}

\author[0000-0003-2658-7893]{Francoise Combes}
\affiliation{Observatoire de Paris, LERMA, Collège de France, CNRS, PSL University, Sorbonne University, 75014, Paris\\}

\author{Pascale Jablonka}
\affiliation{Institute of Physics, Laboratory of Astrophysics, Ecole Polytechnique Fédérale de Lausanne (EPFL), Observatoire de Sauverny, CH-1290 Versoix, Switzerland\\}

\author{Yannick Bah\'{e}}
\affiliation{Institute of Physics, Laboratory of Astrophysics, Ecole Polytechnique Fédérale de Lausanne (EPFL), Observatoire de Sauverny, CH-1290 Versoix, Switzerland\\}

\author[0000-0002-6220-9104]{Gabriella De Lucia}
\affiliation{INAF - Astronomical Observatory of Trieste, via G.B. Tiepolo 11, I-34143 Trieste, Italy\\}

\author{Vandana Desai}
\affiliation{IRSA, California Institute of Technology, MS 220-6, Pasadena, CA 91125, USA\\}

\author[0000-0002-3144-2501]{Rebecca A. Koopmann}
\affiliation{Department of Physics \& Astronomy, Union College, Schenectady, NY, 12308, USA\\}

\author{Dara Norman}
\affiliation{National Optical Astronomy Observatory, 950N Cherry Avenue, Tucson, AZ 85750\\}

\author{Melinda Townsend}
\affiliation{Department of Physics and Astronomy, Siena College, 515 Loudon Road, Loudonville, NY 12211, USA\\}

\author{Dennis Zaritsky}
\affiliation{Steward Observatory, University of Arizona, 933 North Cherry Avenue, Tucson, AZ 85721-0065, USA\\}

\date{\today}

\begin{abstract}
\noindent 
Recent theoretical work and targeted observational studies suggest that filaments are sites of galaxy preprocessing. The aim of the WISESize project is to directly probe galaxies over the full range of environments to quantify and characterize extrinsic galaxy quenching in the local Universe. In this paper, we use GALFIT to measure the infrared 12$\micron$ ($R_{12}$) and 3.4$\micron$ ($R_{3.4}$) effective radii of 603 late-type galaxies in and surrounding the Virgo cluster. We find that Virgo cluster galaxies show smaller star-forming disks relative to their field counterparts at the $2.5\sigma$ level, while filament galaxies show smaller star-forming disks to almost $1.5\sigma$. Our data, therefore, show that cluster galaxies experience significant effects on their star-forming disks prior to their final quenching period. There is also tentative support for the hypothesis that galaxies are preprocessed in filamentary regions surrounding clusters. On the other hand, galaxies belonging to rich groups and poor groups do not differ significantly from those in the field. We additionally find hints of a positive correlation between stellar mass and size ratio for both rich group and filament galaxies, though the uncertainties on these data are consistent with no correlation. We compare our size measurements with the predictions from two variants of a state-of-the-art semi-analytic model (SAM), one which includes starvation and the other incorporating both starvation and ram-pressure stripping (RPS). Our data appear to disfavor the SAM, which includes RPS for the rich group, filament, and cluster samples, which contributes to improved constraints for general models of galaxy quenching.

\end{abstract}


\section{Introduction}

The matter distribution of the Universe on scales of tens of megaparsecs (Mpc) is characterized by regions of varying density that form the cosmic web. This structure is composed of interconnected filamentary networks funneling matter into dense nodes of galaxy groups and clusters, as evidenced by both simulations 
\citep[e.g.][]{sousbie2011A,bahe2017,kuchner2022} and observations \citep[e.g.][]{delapparent1986,bond1996,tifft1976,joeveer1978,tully2014}. The dynamic nature of gravitational attraction ensures that galaxies do not settle into any one environment: fields are increasingly emptied as higher density regions pull matter through ever-narrowing filaments into the outskirts of clusters, and eventually nearer to the cluster cores.

Numerous studies have also indicated that galaxy properties vary with local density. For instance, galaxies in clusters or rich groups tend toward lower star formation rates \citep[e.g.][]{balogh1997,vulcani2010}, smaller star-forming disks \citep[e.g.][]{koopmann2004,schaefer2017,finn2018}, lower gas contents \citep[e.g.][]{boselli2014,chung2009,scott2013}, and earlier type morphologies \citep[e.g.][]{dressler1980,postman1984,goto2003} as compared with the their isolated field counterparts at a given stellar mass. Complicating matters, these same characteristics are also correlated with intrinsic galaxy properties, such as stellar mass \citep[e.g.][]{kennicutt1998,kauffmann2003,vulcani2010,boselli2014,saintonage2017}. As such, nature and nurture conspire to simultaneously drive these galaxy evolution trends, with a varying efficiency depending on galaxy type and environment \citep[e.g.][]{kauffmann2004,peng2010,delucia2012}. Environmental processes range from mergers between galaxies; to the hydrodynamic removal of gas from within a galaxy known as ``ram-pressure stripping" 
\citep[][]{gunn1972,quilis2000,tonnesen2019}; to the decoupling of a galaxy from hot gas halo as it is accreted onto a larger system, called ``strangulation" \citep{larson1980}. Each of these mechanisms has a different effect on the distribution of star formation within galaxies and so is identifiable through unique spatial signatures \citep[e.g.][]{quilis2000,tonnesen2019,kawata2008,vulcani2022}. Ram-pressure stripping, for example, will preferentially remove gas from the outskirts of an infalling galaxy and may induce additional star formation around its center \citep[][]{fujita1999,koopmann2004,fosseti2018,tonnesen2019}, while ``strangulation" primarily affects the galaxy's hot gas content by suppressing gas cooling \citep{larson1980}.

Clusters are the most massive gravitationally-bound structures and are therefore the most extreme sites of environmental processing. And yet, observational studies beyond the cluster outskirts have found that some form of environmental processing occurs before the galaxies reach the cluster, presumably while these galaxies are traveling along filaments \citep[e.g.][]{poggianti1999,lewis2002,gomez2003,cortese2006,bahe2013,laigle2018,kraljic2018,sarron2019}; and early observational studies of filaments have found that they host galaxies with gas contents and morphologies lying between cluster and field galaxies \citep[e.g.][]{castignanivirgo1, castignani2022}. 
Indeed, simulations reveal that cosmic filaments are suffused with gas that can produce effects on their constituent galaxies that quench (or suppress) their star formation rates \citep[e.g.][]{bahe2017,salerno2022}. This full range of environments, however, is not widely used. A more common approach to defining environments has instead been a simplified field/group/cluster scheme, in part because the fraction of quenched galaxies in clusters is noticeably higher than in more isolated regions \citep[e.g.][]{poggianti2006} and in part because filaments are difficult to robustly identify without highly complete catalogs with precise redshift measurements \citep[e.g.][]{castignanivirgo1}.


We know that environment affects galaxy evolution, and that the cosmic web plays host to these effects. However, we only have a preliminary view of where and how star formation is modified within galaxies and how the changes depend on where the galaxies are located within the cosmic web. Observationally, \citet{finn2018} show that the spatial extent of galaxies' obscured star formation, normalized to the spatial extent of their stars, depends on bulge size for late-type galaxies; but even when controlling for the bulge-to-total ratio, this spatial extent of star formation also varies inversely with environment density \citep{finn2018}. That is, there is a supposed link both between the spatial distribution of star formation and both galaxy morphology and environment density, complicating efforts to isolate environmental effects on star formation. Meanwhile, various computational studies indicate that galaxies do experience preprocessing as they travel through filaments, but the predictions differ regarding whether these structures enhance 
\citep[e.g.][]{enhance1,enhance2} or deplete \citep[e.g.][]{deplete1,deplete2} the galaxies' gas reservoir as they funnel into the larger cluster nodes. There are also such studies which report both phenomena depending on a galaxy's proximity to the host cluster \citep[e.g.][]{kotecha2022} or on the overall structure of the filaments (Bahé et al., in prep.). Specifically, Bahé et al. used simulations of areas surrounding various well-resolved clusters to show that even after controlling for stellar mass, gas accretion rates of galaxies in the densest connecting filaments are suppressed, while galaxies occupying the more sparse filaments exhibit elevated accretion rates. These conflicting reports clearly speak to the need for additional investigation of filament effects, as well as of the link between star formation and galaxy morphology.

If we are to understand how galaxies transform, regardless of their place in the cosmic web, then we must use a more nuanced range of environments in a large statistical sample in order to disentangle these various processes affecting star formation rate. The goal of the WISESize project is to combine multi-wavelength diagnostics of spatially-resolved star formation rates with a large sample spanning a more complete environment spectrum. To that end, this project constitutes an investigation of galaxies within and surrounding the Virgo cluster, with the catalog including galaxies within a radius of 24 Mpc relative to its center. We report on parametric measurements of the spatial extent of active star formation in these galaxies using existing imaging of warm dust emission from the WISE satellite (\cite{wright2010}; 12$\micron$), normalized to the size of the galaxies' stellar disks (WISE 3.4 $\micron$). This ratio, $R_{12}/R_{3.4}$, acts as a direct measure of the gas distribution in galaxies with respect to the underlying stellar population. We can then identify quenching mechanisms in different regimes, since the anticipated hydrodynamic processes affect the radial distribution of the star formation differently. Comparing the size ratios of galaxies in different environments, while simultaneously controlling for stellar mass and morphology, allows us to quantify the effect of the full range of environments on star formation.


\S\ref{sec:data} presents the Virgo galaxy sample, the environments of the galaxies, and their 3.4$\micron$ and 12$\micron$ imaging; as well as our applied quality and morphological cuts to ensure that the measurements are as robust as possible. We present our size measurement techniques in \S\ref{sec:methods} and our results in \S\ref{sec:results}, including internally consistent star formation rates (SFR) and stellar mass estimates, a comparison of size ratios with respect to SFR and stellar mass, and median size ratios versus environment and stellar masses. \S\ref{sec:discussion} reviews how these results compare with previous observational and theoretical studies. We also include size ratios plotted atop semi-analytic model (SAM) predictions from the GAlaxy Evolution and Assembly \citep[GAEA]{xie2020} in order to compare our data with theory. Finally, we summarize findings in \S\ref{sec:conclusion} and present our future directions.


\section{Sample Selection \& Data}
\label{sec:data}
The Virgo Filament Survey (VFS) is a collaborative effort to characterize the galaxies within 12 projected virial radii (or $\sim$24 Mpc) of the Virgo cluster center (187.70$^{\circ}$ 12.34$^{\circ}$, J2000), the nearest cluster in the local Universe. The VFS galaxies span a 2D projection of 100$^{\circ}$ $<$ RA $<$ 280$^{\circ}$ and -35$^{\circ}$ $<$ DEC $<$ 75$^{\circ}$ (J2000) and lie within a heliocentric velocity range of 500 $< v_r <$ 3300 km/s. This lower $v_r$ bound limits stars and galactic contamination while the upper bound ensures that filament systems are fully included, as many are situated behind Virgo \citep{mei2007}. The final result is a catalog containing 6780 sources, described in detail in \cite{castignani2022}. 

\subsection{Environment of galaxies}
\label{sec:env}

An exhaustive treatment of our heterogeneous environment classification techniques for the full VFS sample is presented in \citet{castignani2022}. This includes a correction for cosmic flow velocity from in \citet{mould2000} and direct distance measurements from \citet{Steer2017}. This catalog assigns galaxies as belonging to various environments: cluster, rich group, poor group, filament, and field. Galaxies may belong to multiple environments at once (e.g., filament and group), pure field and cluster galaxies notwithstanding. 

Beginning with group membership, \citet{castignani2022} matched each galaxy to the environmental catalog from \citet{kourkchi2017}, which classified galaxy groups in the local Universe (with $v_r <$ 3500 km/s), and then mapped where these groups or isolated galaxies resided relative to the surrounding environmental architecture. Adopting the \citet{kourkchi2017} definition, we define rich groups as containing 5 or more members while poor groups contain $< 5$ members. The cluster environment is defined such that members must be within 3.6 h$^{-1}$ Mpc (where $h$ = 0.74) from the Virgo center coordinates in the 3D Super Galactic frame \citep{castignani2022}. Filamentary membership is based on previously-established structures in the Northern Hemisphere from \cite{kim2016} and \cite{tully1982}, as well as original identification of five supplementary filaments. These five filaments were found using high density contrast relative to the surrounding field as detailed by visual inspection in 3-dimensional supergalactic coordinates, and the authors flagged galaxies within 2 Mpc (in 3D supergalactic space) of a given spine as being part of that filament \citep{castignani2022}. Pure field galaxies, lastly, are simply those which are not flagged as belonging to any of the above environments. 





\subsection{Imaging Data}
\subsubsection{Optical \textit{grz} imaging}
\label{sec:optical}

We use optical \textit{grz} imaging from the DESI Legacy Imaging Surveys \citep[][Data Release 9]{dey2019} for our SED fitting (see \S\ref{sec:magphys}) and the \textit{r}-band specifically to provide position angles, axis ratios, and complementary measures of the spatial extent of stellar disks. These surveys catalogued the northern hemisphere over 14000 square degrees using the \textit{grz} optical bands in conjunction with photometry of mid-infrared bands from the Wide-Field Infrared Survey Explorer satellite (WISE). The Legacy Survey team combined data products from the Blanco telescope at the Cerro Tololo Inter-American Observatory (The Dark Energy Camera Legacy Survey, DECaLS); the Mayall Telescope at the Kitt Peak National Observatory (The Mayall z-band Legacy Survey, MzLS); and the University of Arizona Steward Observatory 2.3m Bart Bok Telescope, also at Kitt Peak National Observatory (Beijing-Arizona Sky Survey, BASS). These data are available at \href{https://www.legacysurvey.org}{https://www.legacysurvey.org}.

\subsubsection{IR imaging}
\label{sec:irimaging}
The WISE satellite imaged the sky in four mid-infrared wavelength channels: 3.4$\micron$, 4.6$\micron$, 12$\micron$, and 22$\micron$ \citep{wright2010}, all of which are publicly accessible. Our infrared data in particular are from unWISE as provided by the Legacy Survey \citep{lang2014}, which reverses the blurring of the original WISE coadd images. We use all four WISE bands for the SED fitting routine, in addition to the \textit{grz} bands mentioned previously and \textit{NUV} and \textit{FUV} bands from GALEX \citep{galex2005}, and the data for which can be found in MAST in the GALEX/MCAT High Level Science Product page \citep{10.17909/T9H59D}. Our primary data for modelling the spatial extent of the stars in the VFS sample is from the 3.4$\micron$ channel, while the 12$\micron$ images are to find the spatial extent of the obscured star formation. The image resolution at both of these wavelengths is sufficient for measuring the relevant spatial profiles. 
While examining the 12$\micron$ images, we discovered that the publicly available unWISE images \citep{lang2014} have systematically oversubtracted the sky background for large galaxies. We corrected this issue and describe the process in Appendix \ref{sec:Appendixbackground}. 



\subsubsection{Stellar Masses \& SFRs}
\label{sec:magphys}
We use the Multi-wavelength Analysis of Galaxy Physical Properties (MAGPHYS; \cite{magphys2008}) to determine stellar masses and star formation rates from our full galaxy sample. We construct our observed SEDs using custom elliptical aperture photometry of large-area public surveys that span the UV \citep[GALEX;][]{GildePaz2007}, optical \citep[$Legacy \ Survey$;][]{dey2019}, and infrared \citep[WISE;][]{wright2010}. The procedure for measuring the photometry and masking contaminants is described in detail in \citet{moustakas2023}. Specifically, our fluxes are measured within a fixed elliptical aperture whose semi-major axis is 1.5 times the estimated size of the galaxy based on the second moment of the light distribution (after subtracting stars and masking out surrounding galaxies in the image). We do not attempt to correct the aperture fluxes to total fluxes. However, using a curve-of-growth analysis, we estimate that the correction would affect the stellar masses by $<20$\%. 

To correct for galactic extinction, we use the reddening values from \citet{Schlegel1998} and follow the Legacy Survey's procedure to transform to the $grz$ and WISE filters. We transform $E(B-V)$ to extinction in the GALEX FUV and NUV filters using the transformations from \citet{Wyder2007}. 

Some of our images have saturated \textit{grz} pixels or a primarily optically-prominent foreground star, which could result in unreliable MAGPHYS results. For these cases, we calculate alternate SFR and stellar mass values based on infrared color-based conversions. For the SFRs, we use the corrected NUV equation from \cite{kennicutt2012}:
\begin{align*}
 \log_{10}(\text{SFR}_{\text{NUV, corr}}) = \log_{10}(\nu_{NUV} L_{\nu_{\text{NUV, corr}}}),
\end{align*}
where $\nu_{\text{NUV}} L_{\nu_{\text{NUV, corr}}} = \nu_{\text{NUV}} L_{\nu_{\text{NUV}}} + 2.26 \nu_{22} L_{\nu_{22}},$ $\nu L_\nu$ and is the luminosity density in erg s$^{-1}$. For stellar masses, \citet{jarrett2023} give two equations based either strictly on $L_{3.4}$ or a combination of $L_{3.4}$ and C$_{12}$ (the W1 - W2 color in Vega magnitudes). We choose the latter due to its being constrained by three variables versus one. This equation is given as
\begin{align*}
 \log_{10}(M_*) = A_0 + A_1 C_{12} + \log_{10}(L_{W1}),
\end{align*}
with $L_{W1}[L_\odot] = 10^{-0.4(M_{W1} - M_\odot)}$, $A_0$ = -0.376, and $A_1$ = -1.053 \citep{jarrett2023}. We use a value of $M_\odot = 3.26$ for the absolute magnitude of the Sun in Vega \citep{jarrett2011}. To convert photometric flux to magnitude, we calculate distances using cosmic recession velocities from \citet{castignani2022}.

We then isolate unsaturated galaxies and fit a line to both the log(SFR$_{\text{MP}}$) vs. log(SFR$_{\text{NUV, corr}}$) and log(M$_{*_{\text{MP}}}$) vs. log(M$_{*_{W1}}$) plots, where MP denotes MAGPHYS values. The fitted relationships are as follows: 
\begin{align*}
 \log_{10}(M_{*_{\text{MP}}}) = m_1 \log_{10}(M_{*_{W1}}) + b_1 \\
 \log_{10}(\text{SFR}_{\text{MP}}) = m_2 \log_{10}(\text{SFR}_{\text{NUV, corr}}) + b_2, 
\end{align*}

where $m_1 = 0.824 \pm 0.010$, $b_1 = 1.397 \pm 0.088$, $m_2 = 1.133 \pm 0.014$, and $b_2 = -0.075 \pm 0.020$. For the galaxies flagged as having any of the above photometric problems, we replace the spurious MAGPHYS results with these derived alternate values from their color-based SFR and stellar mass results.





\subsection{Sample Selection for Spatial Profile Fitting}
\label{sec:trim}
While the VFS contains 6780 galaxies, not every one is suitable in the WISE bands for parametric disk size measurements. We therefore apply multiple quality cuts. \cite{vanderwel2012} found that the measurement uncertainties in a galaxy's structural parameters depend to first order on the signal-to-noise ratio (S/N) integrated over the galaxy, so our foremost cut is on this quantity. Our S/N cut removes galaxies with a ratio $<$10 in the 12$\micron$ band, leaving 2083 galaxies. 



After visual inspection we removed an additional 10 galaxies, as there were bright foreground stars which overlapped the central galaxy or corrupted the image with significant scattered light. Note that a number of central galaxy cutouts also featured secondary or even tertiary galaxies, either as part of the VFS or with a redshift that placed the galaxy beyond the bounds of the VFS. GALFIT is able to perform multiple component fits to the input images, so we marked these postage stamps as containing 2$+$ Sérsic objects to be fit in this way. We fit all galaxies regardless of whether they belong to the VFS, which prevents their flux contributions from being erroneously incorporated into the VFS galaxy models (see \S\ref{sec:galfit} for further details).

We next apply a cut helping control for morphological type using Hyperleda's T-type classification \citep{makarov2014}, motivated by the correlation between B/T and disk size ratios as reported in \citet{finn2018}. A secondary rationale is that early-type galaxies have low star formation rates by definition, so their 12$\micron$ emission may stem from sources unrelated to star formation and therefore do not directly inform us about the modification of the gas and dust distribution in a galaxy \citep{salim2016, kelson2010}. We use a T $\geq$ 0 flag, with T=0 corresponding to S0a galaxies, to select for late-type galaxies in lieu of bulge-to-total ratios as used for the Local Cluster Survey \citep{finn2018}. The sample count following these cuts is now 1653 galaxies.

\subsubsection{Cuts on log{(SFR)}, log{(sSFR)}, and log{(M$_*$})}
\label{ref:logcuts}

\begin{figure}[h]
\includegraphics[scale=0.38]{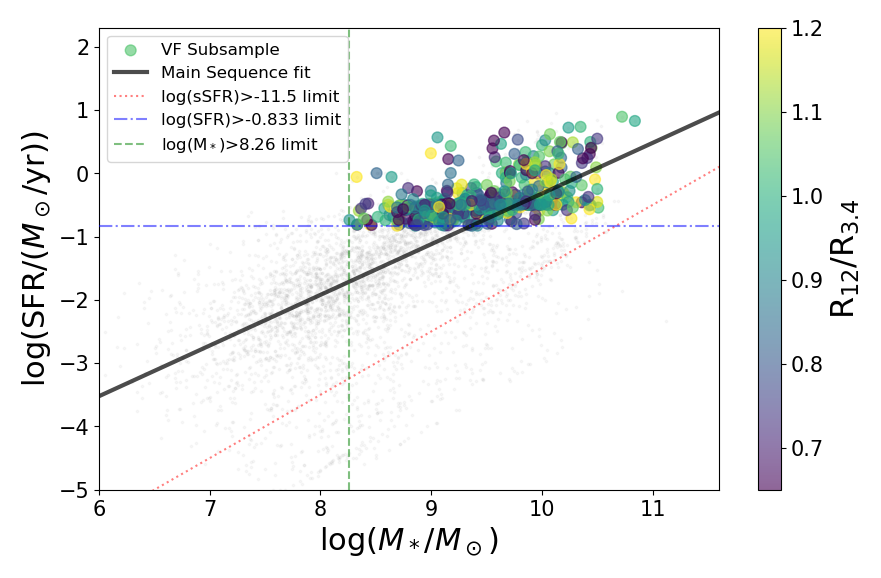}
\centering
\caption{SFR vs. M$_*$ for galaxies in our Virgo sample. The color map depicts the size ratio with values from GALFIT. The light gray background points represent the VFS galaxies. A Main Sequence line is fit to all points with log(sSFR yr$^{-1}$) = $>-11.5$ to illustrate how the sample galaxies compare to the full sample. Galaxies which do not meet the log(sSFR), log(SFR), or log($M_*$) criteria as given in the legend are excluded from our sample. Note that gray points in the upper right quadrant which are not part of the our sample are either flagged as having AGN or (see Section \S\ref{sec:agn}) experienced GALFIT errors and thus do not have robust size ratio measurements.}
\label{fig:sfrmstar}
\end{figure}

Figure \ref{fig:sfrmstar} shows the distribution of our sample on a star formation rate versus stellar mass plot. The color denotes the galaxies' size ratios, R$_{12}$/R$_{3.4}$, which we detail in \S\ref{sec:galfit}, while the gray background points show the full VFS. The origins of the various lines and limits are described below.

We first remove galaxies with log(sSFR yr$^{-1}$) $<-11.5$ as galaxies below this limit are those whose UV and IR emission are likely dominated by sources not associated with star formation \citep{salim2018}. We find the main sequence equation for the Virgo galaxies, 
\begin{equation}
 \log(SFR/M_{\odot}/yr) = 0.80*\log(M_*/M_{\odot}) - 8.32,
\end{equation}
by applying this log(sSFR yr$^{-1}$) $>-11.5$ cut and fitting a straight line to all remaining galaxies in the Virgo sample.

The second cut is $log(\rm{SFR}/\rm{M}_{\odot}/yr) >-1.01$, which we determined by finding the lowest 5\% SFR at the upper 10\% of our galaxies' distance distribution. This cut is to ensure that we are equally complete in SFR over the velocity range of our final sample, $500 < \rm{v}_{\rm{cosmic}} < 3300$ kms$^{-1}$. 

We then determined the stellar mass completeness limit, above which we can detect all galaxies regardless of their $r$-band stellar mass-to-light ratio (${\rm M}_\star/L_r$). We derived the stellar mass completeness limit using a technique adapted from \citet{marchesini2009}, \citet{rudnick2017}, \citet{finn2023}, and \citet{zakharova2024}. The galaxies at our apparent magnitude limit that have the faintest luminosity are those at the high velocity (distance) end of our survey, namely those with $2500 < v/{\rm km/s} < 3500$. We selected all galaxies with $m_r = 16.52 - 17.27$, corresponding to 0.5 and 1.25 mag brighter than the SDSS spectroscopic limit of $m_r = 17.77$. As these galaxies are significantly brighter than the detection limit, we can detect all galaxies with equal completeness, regardless of their ${\rm M}_\star/L_r$. We assume that ${\rm M}_\star/L_r$ does not vary strongly with $m_r$ over this range. Therefore, the distribution of ${\rm M}_\star/L_r$ for this bright subsample should be representative of the intrinsic ${\rm M}_\star/L_r$ distribution at our apparent magnitude limit. We measure the 5\% highest ${\rm M}_\star/L_r$ from this distribution and multiply it by the luminosity that corresponds to $m_r=17.77$ for the most distant galaxies. This yields a stellar mass limit of $\log({\rm M}_\star / {\rm M}_\odot) = 8.26.$ Galaxies at lower stellar masses would only be detectable if they had lower ${\rm M}_\star/L_r$ values. 

Overall, the cuts on SFR, sSFR, and M$_\star$, remove an additional 676 galaxies from our sample. 

\begin{figure}[h]
\includegraphics[scale=0.42]{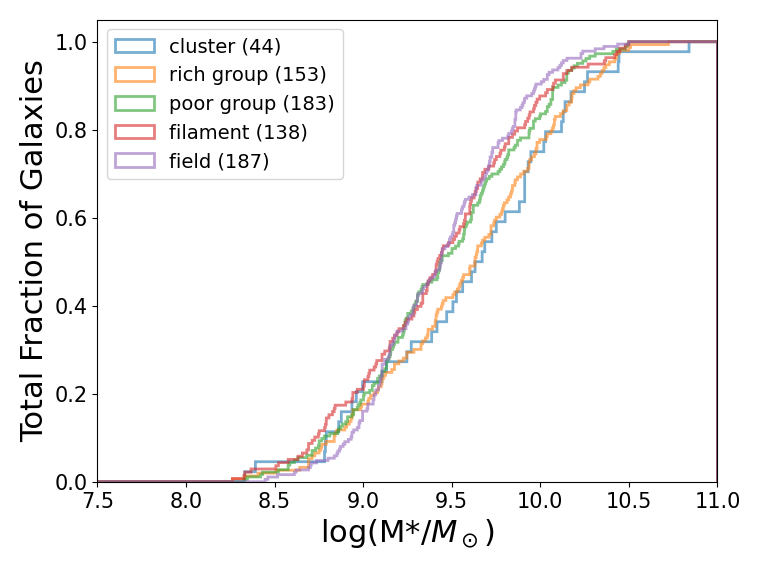}
\centering
\caption{A cumulative histogram plot of the VFS sample comparing stellar mass distributions from MAGPHYS, divided into our five environments. Parenthetical values represent the number of galaxies in each bin. Since multiple galaxies share environment classifications, summing the numbers in the legend will give a different value than our final sample count.}
\label{fig:masscomp}
\end{figure}

Controlling for stellar mass helps isolate environmental effects on size ratio distributions. We present one test of this in Figure \ref{fig:masscomp}, showing a cumulative histogram for the mass distribution in each environmental bin. Only sample galaxies without GALFIT errors are considered (see \S\ref{sec:sizeratenvdensity}). If the distributions are roughly comparable in each environment, differences in mass distribution should not be driving differences in the SFRs. We quantify the similarities in Table \ref{table:ksmass}, where each number represents the Kolmogorov–Smirnov (K-S) test p-value for the galaxy masses in the given two environments. For our sample, the mass distributions are comparable between environments. Indeed, while the rich group and cluster galaxies appear systematically shifted toward lower masses, the K-S cannot rule at the greater than 99\% level that the mass distributions are identical. Nonetheless, to ensure these shifts are not biasing our results, we first mass-matched rich group galaxies to each of the four remaining environments and regenerated our results, then doing the same for cluster galaxies. We find that the uncertainties calculated for both mass-matching procedures are in agreement with the median bootstrapping uncertainties that we find without mass-matching (see \S\ref{sec:sizeratenvdensity}).

\begin{deluxetable}{ccccc}
\tablecaption{K-S tests of stellar mass distributions in different environments \label{table:ksmass}} 
 \tablehead{
 \colhead{Environment} & \colhead{Rich Group} & \colhead{Poor Group} & \colhead{Filament} & \colhead{Field}
 }
\startdata
 Cluster & 0.9684 & 0.2722 & 0.1520 & 0.0372 \\
 Rich Group & & 0.0734 & 0.0383 & 0.0082 \\
 Poor Group & & & 0.8695 & 0.3363 \\
 Filament & & & & 0.2210 \\
\enddata
\tablecomments{The values in this table are p-values that denote the probability that the mass distributions are drawn from the same distribution.}
\end{deluxetable}

\subsubsection{Active Galactic Nuclei}
\label{sec:agn}
A final consideration is Active Galactic Nuclei (AGN), which can dominate the galaxy's infrared flux emission. Rather than eliminate AGN altogether, however, we mark suspect galaxies and later assess whether environmental trends differ with respect to their inclusion or exclusion. We adopt two techniques to select AGN, the first being the use of optical emission line ratios if such data are available \citep{kauffmann2003,kewley2004}. In particular, we calculate the galaxy's OIII-H$\beta$ and NII-H$\alpha$ ratios using values from the NASA-Sloan Atlas \citep{blanton2011}. Using the cut from \cite{kauffmann2003} to discriminate ratios indicating star-forming galaxies versus AGN-dominated galaxies, we find 128 sample galaxies with optically-defined AGN.

The second method involves the use of WISE colors. According to \cite{asmus2020}, AGN can be selected with the criteria [3.4$\micron$]-[4.5$\micron$] $>$ 0.65 and [4.5$\micron$]-[12$\micron$] $<$ 4.0 (\cite{asmus2020}; see also \cite{kirkpatrick2015}, \cite{assef2010}). These constraints only flag 24 of the sample galaxies as hosting IR-luminous AGN. 

We find that the statistical significance of our conclusions changes marginally when excluding the 151 AGN defined in these ways, so we elect to remove these galaxies from our sample. 

\subsubsection{Final Sample}

\begin{figure*}[t]
\includegraphics[scale=0.42]{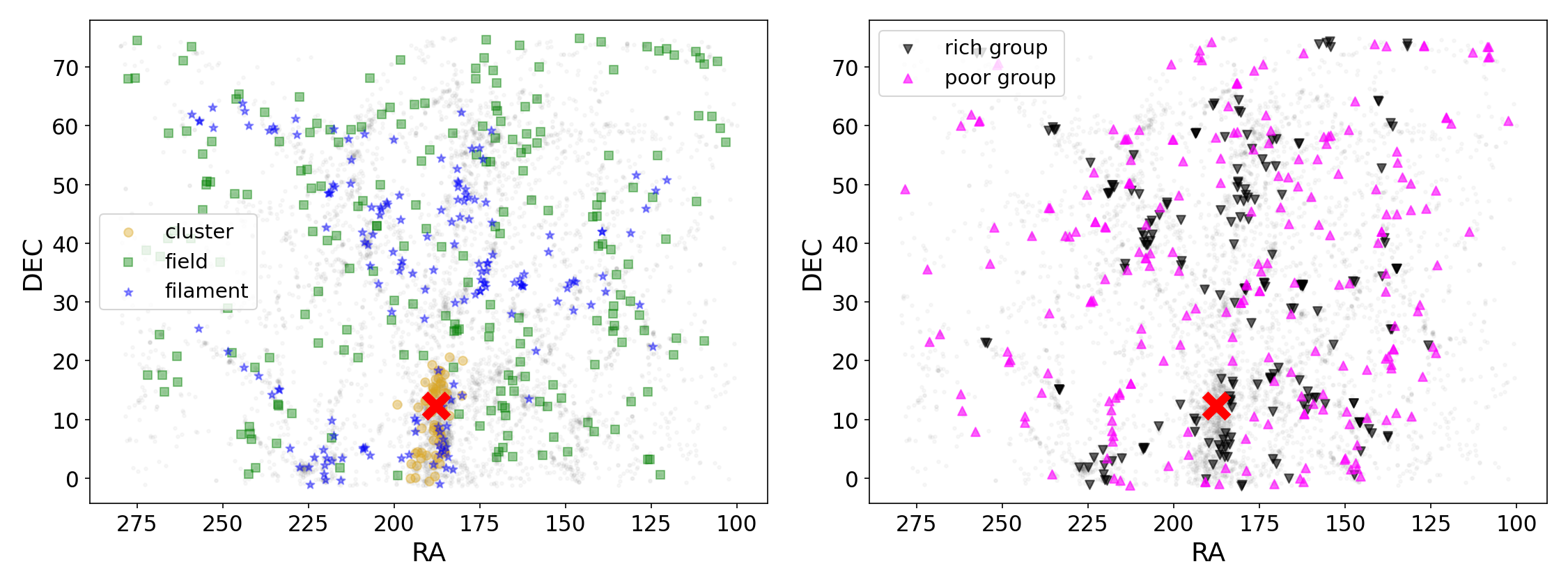}
\centering
\caption{RA-DEC plots of our VFS sample overlaid with environment membership from \citet{castignani2022}. The red X markers represent the Virgo cluster center while the small gray background points show the full 6780 catalog sample for reference.}
\vspace{.5mm}
\label{fig:radec_env}
\end{figure*}

Our model-ready sample contains 997 galaxies. Figure \ref{fig:radec_env} displays these galaxies in RA-DEC space, color-coded according to the environmental flags from \citet{castignani2022}. 
Of these, 7.3\% are cluster members, 25.8\% are in rich groups, 29.4\% are in poor groups, 24.1\% are in filaments, and 31.6\% are in the field. Note that many galaxies share environments (see \S\ref{sec:env}): 5.3\% of galaxies are located in both filaments and poor groups, and 11.4\% belong to both filaments and rich groups.

\section{Methods}
\label{sec:methods}
\subsection{Tracing Stellar Mass and Star Formation with WISE}
We use the WISE W1 images to measure the spatial extent of stellar mass, as this provides a robust, single-wavelength tracer \citep{jarrett2013,mcgaugh2015,durbala2020,jarrett2023}. For the spatial extent of star formation, we use the 12\micron \ emission \citep{calzetti2005,alonso-herrero2006,jarrett2013}. The 12\micron \ band is sensitive to vibrational emissions of polycyclic aromatic hydrocarbons, which are strongly correlated with star-formation rate \citep[e.g.][]{wu2005,maragkoudakis2018,gao2022, whitcomb2023}. Note that we are not extracting an \textit{absolute} star-formation estimate from the 12\micron \ emission, using it instead to indicate where star formation is happening within each galaxy.

\subsection{Structural Parameters from GALFIT}

\label{sec:galfit}

To determine disk sizes of our galaxies, we use parametric fitting of each galaxy cutout with GALFIT as our modeling tool \citep{peng2010}. The fitted parameters include: the central x ($x_c$) and y ($y_c$) pixels of the galaxy, the magnitude (m), the effective radius ($R_e$) enclosing 50\% of the galaxy's total flux, the Sérsic index (n), the position angle (PA), and the axis ratio (B/A). The Sérsic profile \citep{sersic1963} describes the light distribution profile of galaxies as follows: 
\begin{equation}
 I(r) = I_e \exp{\Bigg[-b_n\Bigg(\Big(\frac{r}{R_e}\Big)^{1/n}-1\Bigg)\Bigg]},
\end{equation}
where $I(r)$ is the intensity, $I_e = I(R_e)$ is the intensity at the half-light radius, and $b_n = f(n)$ is a pure function of n and is defined such that the effective radius encloses 50\% the total flux of the profile. The Sérsic index then dictates the shape of the intensity profile. 


We run GALFIT twice, both with and without convolution by the PSF, on each band. For the first pass, we use the following initial conditions: $R_e=25$\arcsec, $n=2$, $B/A=1$, and $PA=0$. We vary the initial guess for the magnitude of the galaxy with the bandpass, and we select the initial value from the peak of the histogram of total magnitudes from our aperture photometry (\S\ref{sec:magphys}). 
As we are working with high SNR images, the results for the first, unconvolved model do not depend strongly on the initial conditions. For the convolved run, we use the unconvolved results as input guesses, as doing so both reduces the total computational time and lowers the amount of numerical errors in our models. Because of the lower spatial resolution of the WISE imaging, we hold B/A and PA fixed at the best-fit $r$-band values and only fit the magnitude, effective radius, and Sersic index for W1 and W3.


Accurate PSF characterization is critical for modeling sources with GALFIT, especially where the PSF blurring is significant. Our infrared images originate from the unWISE catalog, with variable PSFs depending on the declination coordinates of the source \citep{schlafly2019}. We matched the coordinates of each VF galaxy's RA and DEC in our sample with the nearest coadd image's center coordinates, and in this way assigned each galaxy with its quasi-unique PSF.\footnote{See https://github.com/legacysurvey/unwise\_psf} 

For each galaxy cutout, we create a mask using the segmentation image produced by Source Extractor package \citep{sextractor1996}. We additionally mask GAIA stars\footnote{https://www.cosmos.esa.int/gaia} with significant proper motions (SNR of proper motion $> 5$). We visually inspect each mask, as Source Extractor occasionally shreds well-resolved massive galaxies or misses certain background objects, and modify them accordingly. To limit shredding, we enclose our source galaxies with an ellipse 3x its R25 value and forbid Source Extractor from applying its own masking routine to this area (though Gaia sources within this aperture will still be masked).

\begin{figure*}[t]
\includegraphics[scale=0.45]{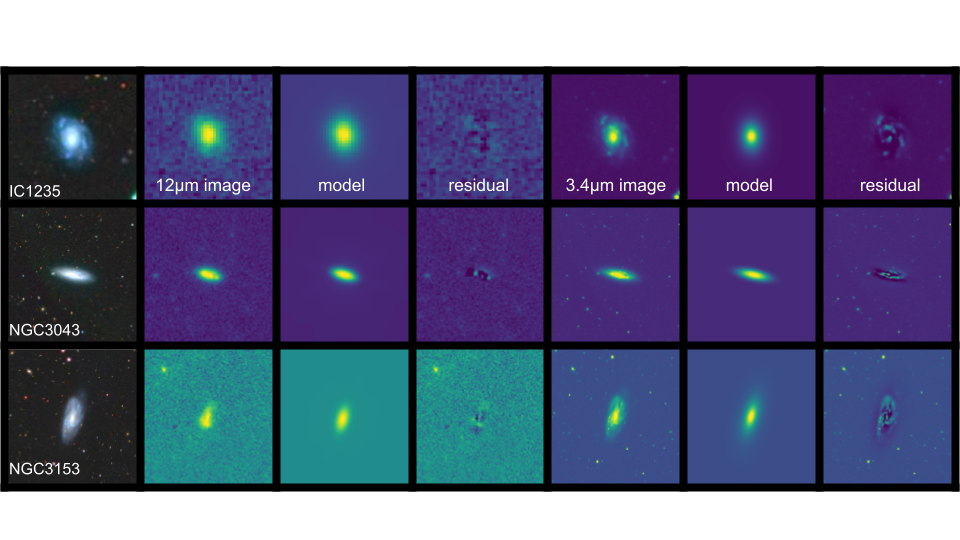}
\centering
\caption{Three example galaxies on which we ran GALFIT, for both the 12$\micron$ and 3.4$\micron$ wavelengths. Each best-fit model incorporated convolution using a custom PSF. The leftmost column is the Legacy $grz$ image for each galaxy, provided as a reference. The residual columns are the result of evaluating [image]-[model], with a pixel color stretch according to the original image.}
\label{fig:exgalfit}
\end{figure*}

In Figure \ref{fig:exgalfit} we present examples of single-Sérsic model fitting for three galaxies using GALFIT, including the models for both the 12$\micron$ case and the 3.4$\micron$ case. From the models' output parameters we record the $R_e$ for each galaxy and calculate the size ratio, $R_{12}/R_{3.4}$. These values will serve as our data for the following sections.



\subsection{Filtering GALFIT Errors}
We exclude galaxies for which GALFIT assigns an error flag, indicating an unreliable output model due typically to non-convergent parameters. Among these galaxies are those with intrinsically small sizes, therefore resembling the PSF shape once convolved. This result typically activates the error flag for the 12$\micron$ convolution cases, as it enables the galaxy model's effective radius to become smaller than the WISE resolution limit in that band and reach sizes even less than a pixel (whereas, for instance, the $r$-band resolution limit is much smaller and therefore does not yield as many error flags). We then remove models that were either poorly constrained, such as when the generated R$_e$ error is greater than twice the output R$_e$; or if $n>6$, translating to an unphysical galaxy flux profile. This vetting process removes 300 total galaxies. This final sample has 603 galaxies in total. Broken down by environment, 44 galaxies are in clusters, 153 in rich groups, 183 in poor groups, 138 in filaments, and 187 in the field. The galaxies with GALFIT errors comprise $\sim 12-17\%$ in each environment category.



\section{Results}
\label{sec:results}

\subsection{R$_{12}$ and R$_{3.4}$ Comparison}

\begin{figure*}[t]
\includegraphics[scale=0.30]{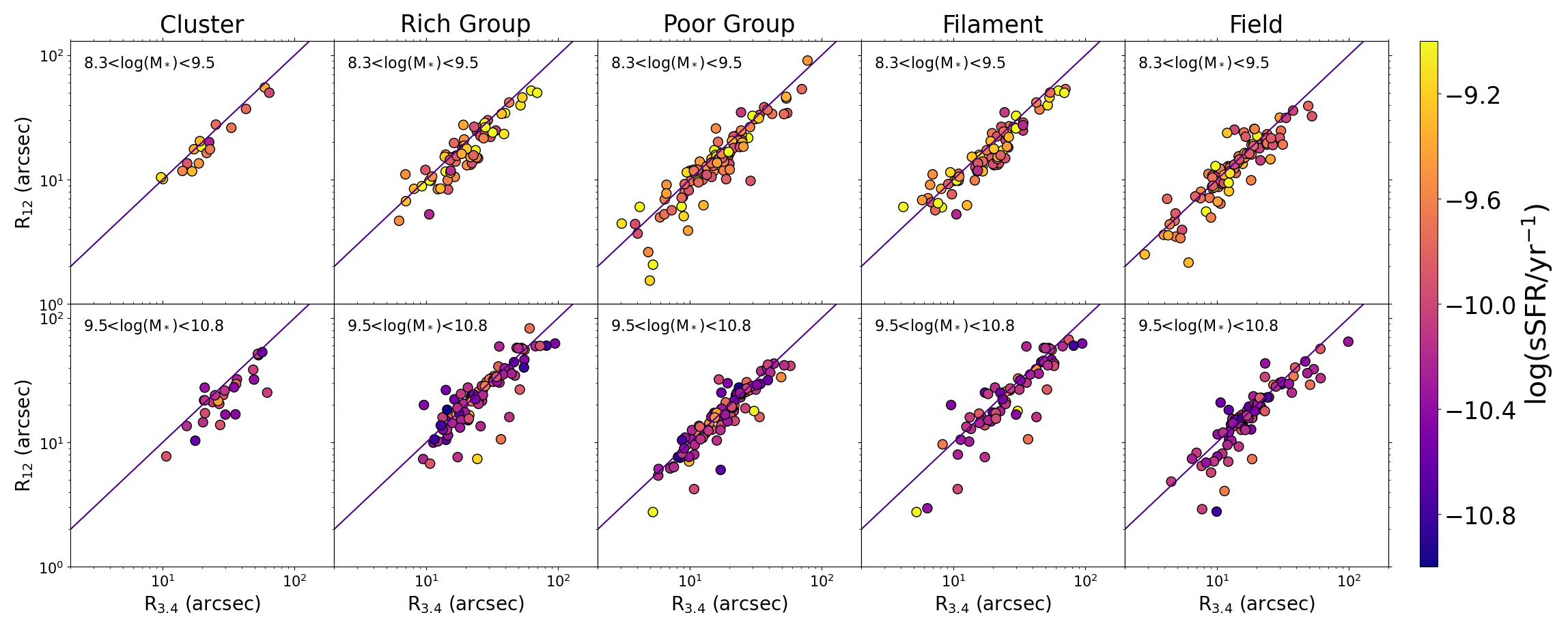}
\centering
\caption{12$\micron$ vs. 3.4$\micron$ effective radii, separated into five environment columns and two mass rows: ($8.3 < \log{M_*} < 9.5$) and ($9.5 < \log{M_*} < 10.8$). Points are color-coded according to sSFR. Each row corresponds to galaxies residing in different environments. 1-to-1 lines are included for reference.}
\label{fig:r12_rstar_all}
\end{figure*}

We first compare the 12$\micron$ R$_e$ to the 3.4$\micron$ R$_e$, as shown in Figure \ref{fig:r12_rstar_all}. The columns of the grid divide the plots into our five galaxy environments: cluster, filament and groups, and field. The rows are either for galaxies in our sample with stellar masses $9.5<\log{M_*}$ (top) or $\log{M_*}>9.5$ (bottom). The colorbars all span the same $\log{sSFR}$ range. The overall trend in each panel is that most of the galaxies follow the 1-to-1 line. Galaxies in the lowest mass range have a relatively larger fraction of the highest sSFR values according to the overall color distributions of each row, though none of the panels show a clear relationship between sSFR and the disk sizes. When comparing the panels both row-wise and column-wise, we also do not find any strong evidence for correlations between the amount of scatter about the 1-to-1 lines and environment density. Fascinatingly, the panels with the most scatter are the field (log(M$_*$)$<9.5$; 0.98) and the cluster (log(M$_*$)$>9.5$; 0.52).


\begin{figure}[h!]
\includegraphics[scale=0.37]{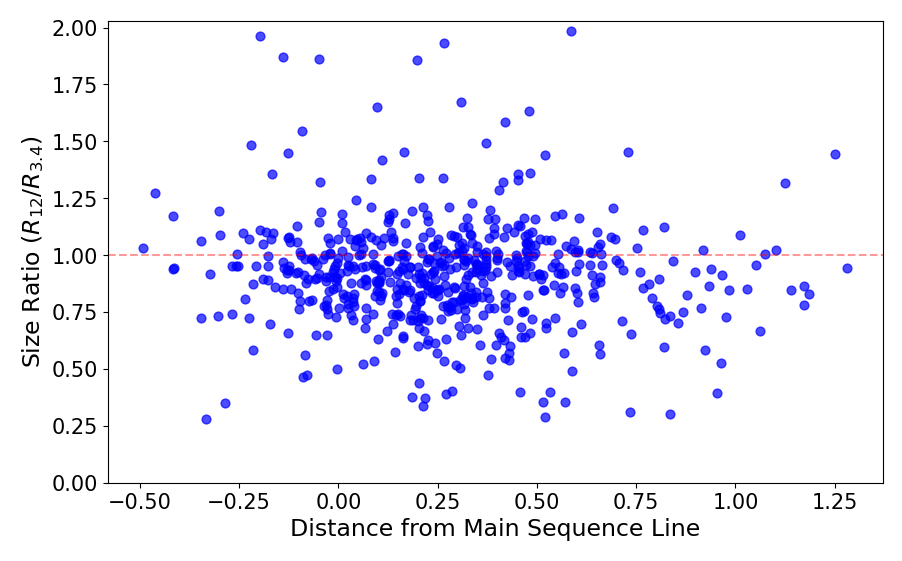}
\centering
\caption{Size ratio as a function of galaxy Main Sequence perpendicular distance. All featured galaxies are from the VFS sample after GALFIT quality cuts. The red dashed line outlines the y-value at which the effective radii values are equal ($R_{12}$ = $R_{3.4}$). The few galaxies with a size ratio $>2$ are excluded for clarity but are still included in our calculations.}
\label{fig:msdist}
\end{figure}

Using the SFR vs. stellar mass main sequence (MS) relation, we also plot the GALFIT size ratios as a function of their perpendicular distance to the MS line in Figure \ref{fig:msdist}. Around 70\% of the galaxies lie below the horizontal line marking the case where $R_{12}$/$R_{3.4}=1$. Eighteen galaxies have size ratios larger than 1.5, of which five are greater than 2. 83\% of the sample galaxies lie within 0.5 dex of the $R_{12}/R_{3.4}$ line. Using a Spearman rank test we find that there is no significant correlation between the size ratio and the offset from the main sequence (Spearman rank coefficient=-0.041, $p=0.407$). We also do not find a correlation between this main sequence offset and environment membership.

\subsection{Size Ratios and Environment Density}
\label{sec:sizeratenvdensity}
\par
\begin{figure}[h]
\includegraphics[scale=0.343]{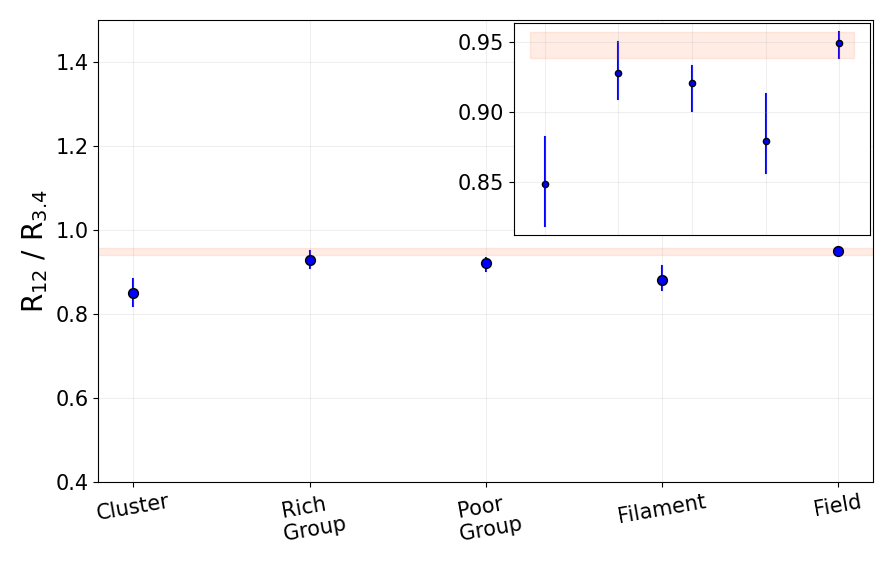}
\centering
\caption{Median size ratios (R$_{12}$/R$_{3.4}$) according to environment bin: cluster, rich group, poor group, filament, and field. Uncertainties represent the 68\% confidence interval on the median found through bootstrapping with 100 iterations to ensure convergence. The shaded region highlights the 68\% spread of the median field ratios, which serves as the control for our study. The upper right plot shows the same data but with a limited y-axis range.}
\label{fig:sizeratenv}
\end{figure}
\par
Figure \ref{fig:sizeratenv} presents the median size ratios divided into five environmental bins. Error bars are derived from bootstrapping iterations and represent the 68\% confidence intervals in the median. The orange shaded region is the confidence interval for the field sample. The main plot's y-axis range reflects the range in Figure \ref{fig:sizerat_pred}, while the upper-right panel is a zoomed-in version to better highlight the median distributions. The overall figure shows two main results. First, we find that the ratio $\rm R_{12}/R_{3.4}$ is less than one in all environments, including the field, implying that the $R_e$ values for dusty star-forming disks (R$_{12}$) are typically smaller than the stellar disks (R$_{3.4}$) for our sample galaxies. Second, $\rm R_{12}/R_{3.4}$ is smaller in the cluster environment, with a median size ratio for the cluster galaxies of $0.834^{+0.047}_{-0.017}$ compared to $0.950^{+0.007}_{-0.011}$ for the field. The poor group and filament medians also differ from the field median by at least a factor of 1$\sigma$. Moreover, while there are significant trends that do exist, the amount of scatter in each environment bin is large. The rich and poor group medians display strong overlap with each other, as do those of the cluster and filament environments. 

\begin{figure}[h]
\includegraphics[scale=0.40]{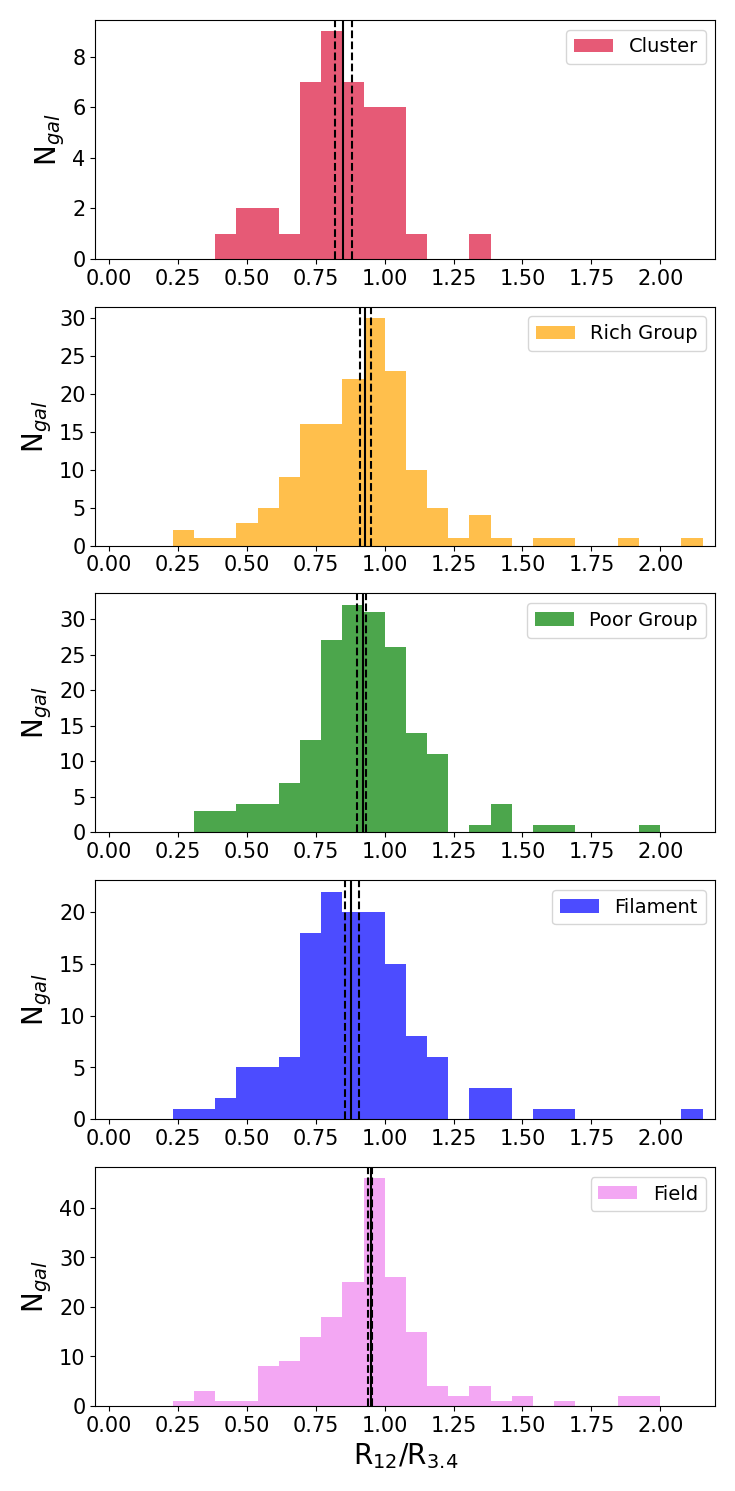}
\centering
\caption{Size ratio distributions for each of the five environments, from most to least dense. The solid vertical lines give the median for each distribution, and the set of dashed vertical lines show the 68\% confidence intervals in the median found using bootstrapping.}
\label{fig:histdist}
\end{figure}

We show the full size distributions as a function of environment in Figure \ref{fig:histdist}. Table \ref{table:ksenv} reports on the p-values for various two-sample tests between these distributions: K-S, Mann-Whitney U (W-M), and Anderson-Darling (A-D). For the field and cluster samples, the three tests respectively yield p-values of 0.0009, 0.0009, and 0.0012. The cluster-field medians expectedly show the most robust difference in sizes, seconded by the field and filament medians (K-S p=0.0032, W-M p=0.0038, A-D p=0.0048). 




Referring to Figures \ref{fig:msdist} and \ref{fig:sfrmstar}, we also do not find a clear dependence of location in the SFR-M$_*$ plane with size ratio. It could be, however, that the smallest size ratios more often than not also have marginally lower SFRs and relatively high stellar masses.

\subsection{Size Ratios and Clustercentric Distances}
\par
\begin{figure*}[t!]
\includegraphics[scale=0.55]{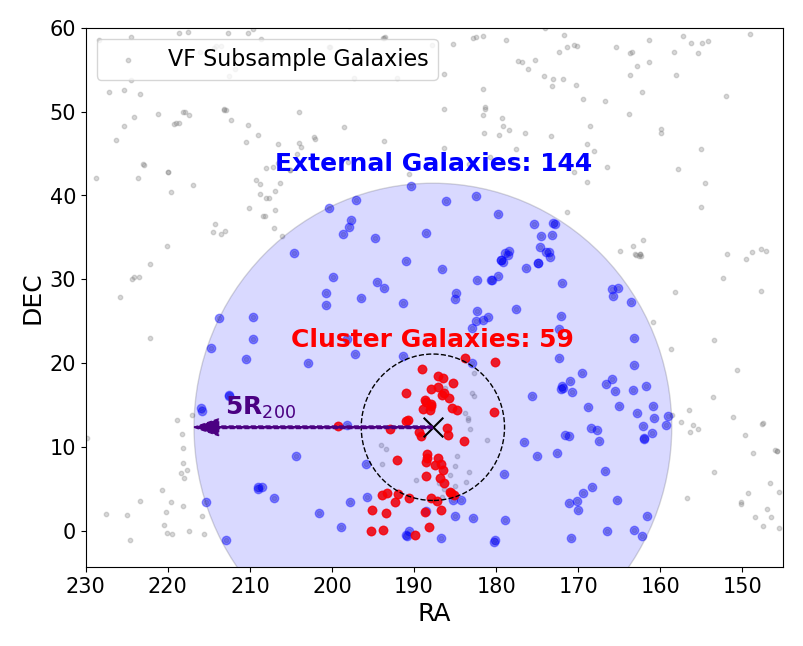}
\centering
\caption{The spatial distribution of our sample divided into cluster and external regions. The external region, denoted by the blue circle, encompasses the radial range between $1.5<R/R_{200}<5$ while the cluster region are galaxies that either at $R<1.5R_{200}$ or are flagged as cluster members according to their phase-space position \citep{castignani2022}. The X represents the Virgo cluster center. The number of galaxies in each set are listed next to the environment label.}
\label{fig:infallcoremap}
\end{figure*}
\par
\begin{deluxetable*}{c ccc ccc ccc ccc}
\tablecaption{K-S, W-M, and A-D tests of size ratios in different environments \label{table:ksenv}}
\tablehead{
\colhead{Environment} & \multicolumn{3}{c}{Rich Group} & \multicolumn{3}{c}{Poor Group} & \multicolumn{3}{c}{Filament} & \multicolumn{3}{c}{Field} \\
& \colhead{K-S} & \colhead{W-M} & \colhead{A-D} & \colhead{K-S} & \colhead{W-M} & \colhead{A-D} & \colhead{K-S} & \colhead{W-M} & \colhead{A-D} & \colhead{K-S} & \colhead{W-M} & \colhead{A-D}
}
\startdata
 Cluster & 0.0247 & 0.0479 & 0.0621 & 0.0423 & 0.0285 & 0.0425 & 0.4385 & 0.3218 & 0.5040 & \textbf{0.0019} & 0.0062 & 0.0078 \\
 Rich Group &&& & 0.6715 & 0.8826 & 0.8746 & 0.2384 & 0.2259 & 0.3669 & 0.4032 & 0.3171 & 0.5991 \\
 Poor Group &&& &&& & 0.2599 & 0.1298 & 0.1801 & 0.1623 & 0.3623 & 0.4310 \\
 Filament &&& &&& &&& & 0.0132 & 0.0229 & 0.0235\\
\enddata
\tablecomments{The values in this table are K-S, W-M, and A-D p-values for the comparisons of the size ratio between different environments. P-values with $\ge 3\sigma$ significance are shown in bold.}
\end{deluxetable*}

An alternative approach to defining galaxy environment is by splitting galaxies according to their proximity to the Virgo cluster center. We define two non-overlapping environment bins: cluster members, which follows the flag criteria given in \cite{castignani2022} \textbf{or} lies within 1.5R$_{200}$ of the cluster center; and external galaxies, which are within two-dimensional 1.5R$_{200}$ and 5R$_{200}$. The Virgo cluster is elongated in DEC (Figure \ref{fig:radec_env}), which motivated the use of the VFS cluster flag rather than restricting core membership to galaxies within strictly 1.5R$_{200}$. The additional criterion of 1.5R$_{200}$ helps to isolate galaxies most likely to be influenced by the cluster environment. The outline of the external and cluster regions is illustrated in Figure \ref{fig:infallcoremap}. These cuts yield totals of 144 external galaxies and 59 cluster galaxies. If we only include galaxies with the \cite{castignani2022} cluster flag, the latter number lowers to around 40.

\begin{figure}[h]
\includegraphics[scale=0.37]{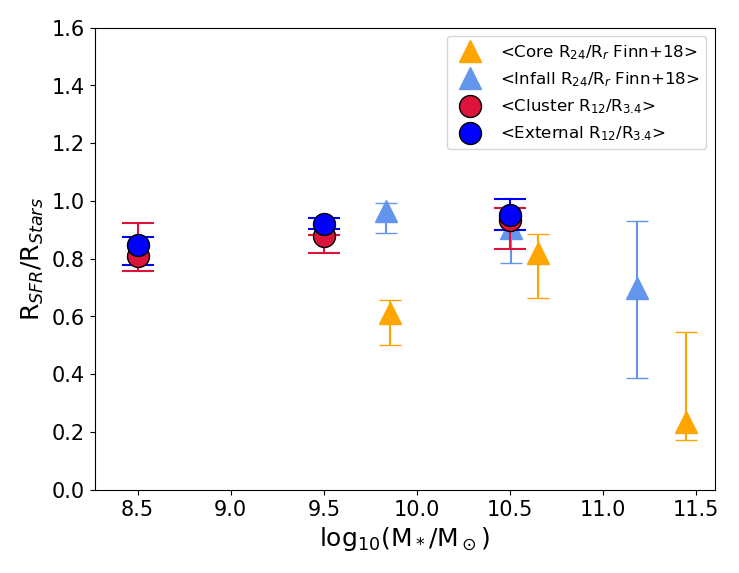}
\centering
\caption{Our median size ratios for three stellar mass bins according to cluster (red circles) or external membership (blue circles), overlayed with the single-component size ratios for core (orange triangles) and infall (light blue triangles) galaxies from \citet{finn2018}. In this case, we define cluster galaxies as those either flagged as cluster members in \citet{castignani2022} \textbf{or} galaxies which lie within 1.5R$_{200}$ of the cluster center according to their 2D projected distance in RA-DEC space. Size ratios from \citet{finn2018} are scaled according to the relationship we found of our GALFIT sizes R$_{r}$/R$_{3.4}$ vs. stellar mass to allow for comparisons between star-forming disks normalized by $r$-band sizes and those normalized by 3.4\micron\  sizes. Error bars originate from bootstrap resampling of the galaxies in each bin and show the 68\% confidence interval for the median points.}
\label{fig:wisesizemass}
\end{figure}

The median size ratios in three bins of stellar mass are compared in Figure \ref{fig:wisesizemass} and do not show significant differences in median size ratio for any individual mass bin, though there appears to be a marginal separation between ``external" and ``cluster" size ratios for the intermediate masses. There is similarly no statistically significant variation in size ratios across our mass bins. There may be too few galaxies to permit this sort of separation according to stellar mass. 


\section{Discussion}
\label{sec:discussion}

\subsection{Size Ratio Model Predictions}
\label{sec:models}

We compare our observations with the theoretical predictions from \citet{xie2020} within the framework of the GAEA\footnote{Information about the GAEA model and selected model predictions can be found at https://sites.google.com/inaf.it/gaea} semi-analytic models. In particular, we plot our data against two models that utilize different prescriptions for the modification of gas in satellite galaxies. The ``GRADHOT" model, which treats gas removal by strangulation \citep{larson1980}, includes a treatment for non-instantaneous stripping of hot gas due to both ram-pressure and tidal stripping \citep[we refer to][for details]{xie2020}\footnote{Tidal stripping would also affect the stars, in an identical way to the gas. While this might occur in the observations, it is not included the GAEA models.}. The ``RPSCOLD" model adds an explicit treatment for the ram-pressure stripping (RPS) of the cold gas within galaxies. This RPS preferentially affects gas in lower surface density regions, thus resulting in outside-in stripping. The GRADHOT and RPSCOLD models successfully reproduce the quenched fraction of central and satellite galaxies at $z=0$ down to $\log_{10}(M_\star/M_\odot) =9.5$ \citep{xie2020}. Additionally, \citet{xie2020} found that the RPS of cold gas has a minor effect on the quenched fractions for massive satellite galaxies for two reasons: massive galaxies have a larger restoring force, so that RPS is less efficient; and massive satellites have spent most of their time as centrals and have been accreted only recently \citep[see e.g][]{delucia2012}, which means that environmental effects have less time to act in general on these galaxies. For low mass satellite galaxies, RPS of the cold gas can increase the quenched fraction and decrease the HI fraction. These two models are among the first to be able to match the quenched fractions reliably in different environments. However, it is impossible from the quenched fraction alone to distinguish between the GRADHOT and RPSCOLD models as they match the quenched fractions equally well.

To build comparison samples from the GAEA models, we measure the half-mass radius for stellar components and the the effective radius of SFR following the method described in \citet{xie2017}. More specifically, we extract the half-mass radius from the total density profile by assuming a Jaffe law \citep{jaffe1983} for the stellar bulge, and an exponential profile for the stellar disk. The scale radius of the disk and effective radius for the bulge are computed within GAEA in a self-consistent way by tracing angular momentum exchanges between different components and following energy conservation arguments \citep{xie2017}. This is similar to our method of fitting a single Sers\'ic profile to the stellar disk. The effective radius of SFR, which is defined as the radius that encompasses half of the total SFR, is extracted from the SFR profile. In the framework of GAEA, SFR is calculated in 21 logarithmically-distributed annuli. At each time step the gas may be truncated by RPS. Following this, the radial arrangement of the gas is redistributed into an exponential using the angular momentum after the stripping and the maximum circular velocity at the time of infall \citep{xie2020}. Outside-in stripping will result in disks with a smaller effective radius for star formation. Because the star formation profile is redistributed to an exponential form, this method is somewhat similar to our observational method of fitting the WISE data with a single S\'ersic profile. 

The different model environments derive from a snapshot of GAEA at z=0. Environments are defined by the halo mass: clusters -- $\log_{10}(M_{\rm halo}/M_\odot) > 14$; rich groups -- $13 < \log_{10}(M_{\rm halo}/M_\odot) < 14$; and field $\log_{10}(M_{\rm halo}/M_\odot) < 12.5$.\footnote{We do not include poor groups as it is not clear what halo mass selection from GAEA would correspond to these systems, which are identified observationally based on their number of member galaxies.} Filaments are identified from the positions of simulated galaxies for the entire simulated GAEA cube by running DisPerSE~\citep{sousbie2011B} with a persistence level 10$\sigma$. Filament galaxies are identified as those in halos with $\log_{10}(M_{\rm halo}/M_\odot) < 14$ (to remove cluster members) and within $2 h^{-1}$Mpc from the nearest filament spine \citep{zakharova2023}. \citet{zakharova2024} explores the effect of peculiar velocities on the identification of filaments and filament galaxies. They conclude that peculiar velocity effects neither alter their results nor significantly impact the DisPerSE classification of filament spines. We apply the same observational limits in $M_\star$, SFR, and sSFR to the model galaxies as we apply to the observed sample.

\begin{figure*}[t]
\includegraphics[scale=0.24]{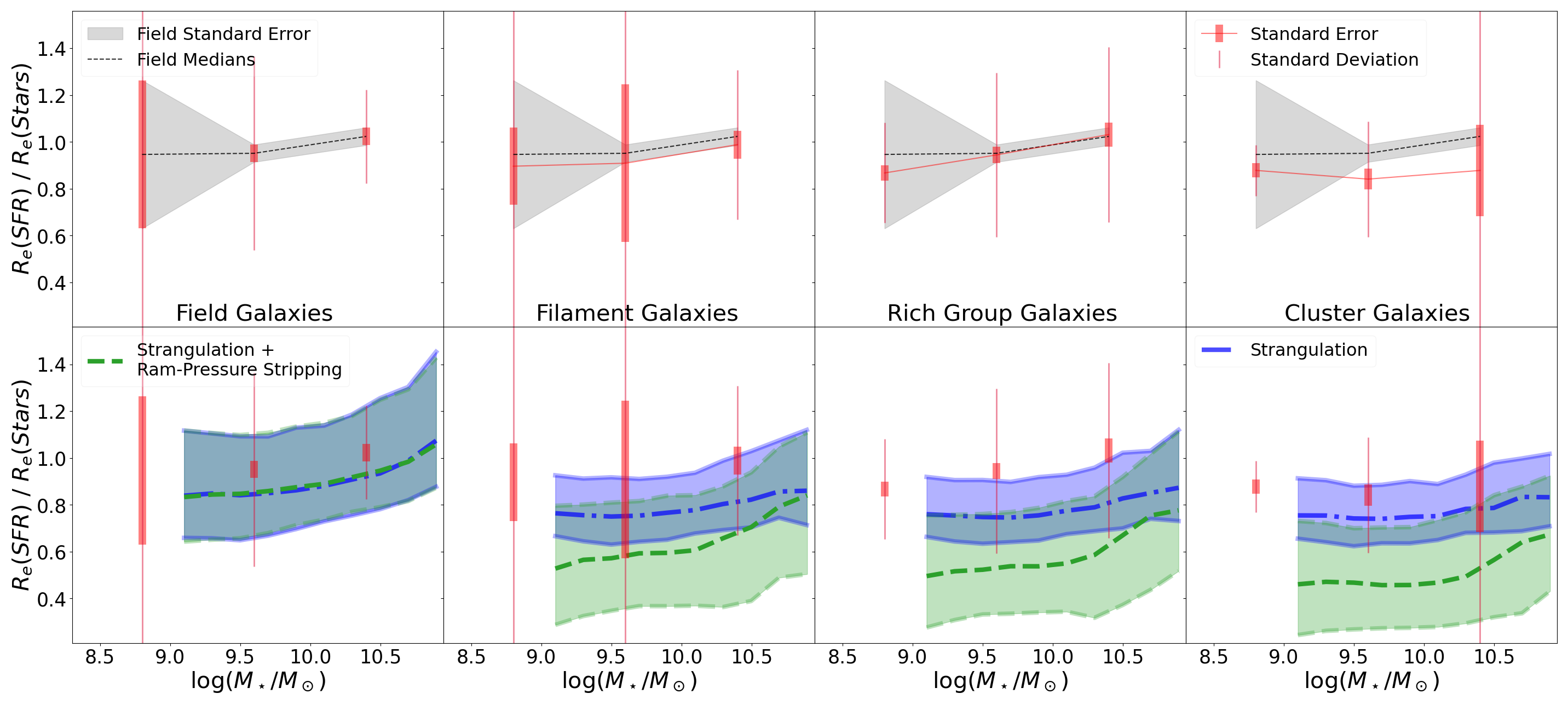}
\centering
\caption{(Top) We isolate four environment regimes (Field, Filament, Rich Group, and Cluster) and divide the galaxies into three stellar mass bins. The thin red lines encompass 68\% of the size ratio data in each mass bin, while the thick red bars trace the standard error on the medians. The gray-shaded region shows the extent of the field standard error, plotted across the three top panels. The dark gray dashed line connects the field points and is shown in each of the remaining panels. (Bottom) Model predictions for two different environmental mechanisms: strangulation (GRADHOT -- blue) and ram-pressure stripping + strangulation (RPSCOLD -- green). The shaded regions for each model encompass 68\% of the model galaxies at each mass. The standard error on the median model size ratio is much smaller, on the order of the thickness of the lines, and for clarity is not included on these plots. The models nearly overlap for the field but are different at a high level of significance in the three other environments. Our data for each environment are reproduced in these bottom panels for comparison.}.
\label{fig:sizerat_pred}
\end{figure*}

In Figure \ref{fig:sizerat_pred}, we compare the GAEA predictions to our observations. Each panel shows the results for one of four environments: field, filaments (including groups within filaments), rich groups, and clusters. The red points show our observations, and the curves show the results from the two different models of environmental processing. The blue curve shows the effects of strangulation (GRADHOT), and the green curve shows the combined effects of strangulation and ram-pressure stripping of the cold gas \citep[RPSCOLD;][]{xie2020}. In the GAEA simulations implemented in the Millennium II simulations, the galaxies are only well resolved down to $\log_{10}(M_\star/M_\odot) =9.0$, which establishes the effective lower-mass limit for the models \citep{xie2020}.
The two predictions for field galaxies (left panel) are nearly identical. This is expected as the different environment processes should not be acting on central galaxies \citep{xie2020}.\footnote{In \citet{xie2020} the two models also do not perfectly overlap in their predictions for the quenched fractions for central galaxies. These slight differences may originate because some of the galaxies that are in halos with $\log_{10}(M/M_\odot) < 12.5$ may actually be satellites.} For the two models, the medians differ at 
a high level of significance given their very small standard error in the median. Their distributions also differ markedly, as shown by the 68\% confidence limits of the distributions shown by the shaded regions. The GRADHOT model extends to somewhat higher size ratios than the RPSCOLD model, while the RPSCOLD model extends to much lower size ratios than the GRADHOT model, as would be expected from the stronger stripping included in the RPSCOLD model. It is somewhat unexpected that the model galaxies undergoing strangulation in the GRADHOT model show a significant reduction in their size ratios compared to field galaxies, as the GRADHOT model does not include any radially dependent removal of gas. One possible explanation is that field galaxies have continued to accrete gas after the galaxies in denser environments have had their accretion shut off. Since accretion at late times in the models is dominated by higher angular momentum gas, it will be deposited at large radii. Thus, the size difference in the GRADHOT model may not be driven by gas stripping, but rather by extra accretion in the field (mostly central) galaxies. Verifying this hypothesis will require a detailed examination of the models that is outside the scope of this paper but will be pursued in a future work.

As shown in Figure \ref{fig:sizerat_pred}, our data in the field are consistent with both models, having only an $\sim 8\%$ larger size ratio. This is itself interesting, as the models have not been calibrated in any way on the size or spatial extent of star formation. Moreover, the way in which the models treat the spatial distribution of star formation, and subsequently measure it, is not perfectly comparable to the data. This supports the applicability of this data-model comparison.
In the group and filament environments, the galaxies have systematically higher size ratios than the models, though they are still broadly consistent with the GRADHOT model. In these environments, the RPSCOLD model appears to be ruled out at modest significance, which is not surprising given that vigorous RPS may not be effective in group and filament environments. In clusters the data are fully consistent with the GRADHOT model and well above the RPSCOLD model. We also perform a Spearman Rank test between the size ratio and stellar mass distributions for each environment, finding a significant correlation with rich groups (p=$5.234 \times 10^{-5}$) and a marginally significant correlation for filament galaxies (p=0.012). 


The preference of our measurements for the strangulation-only models seems to be at odds with the quenching timescale results from \citet{xie2020}, which seem to slightly prefer the RPS models (though only at low masses). It may be that the impact of RPS on gas disk size is therefore over-estimated in \citet{xie2020}, although the quenched fraction is in good agreement with observational measurements. However, we must be cautious with this interpretation as we have a modest sample size and are comparing to only one observed cluster. Measurements of size ratios can nonetheless clearly provide additional discriminatory power to refine the physical processes in the models in a way that cannot be accomplished by only considering quenched fractions. 

Note that the sample extracted from GAEA was drawn from 1/4th the volume of the Millennium II simulation box, which contains many more clusters and also a significant volume that is not in the vicinity of any cluster. This sample is in contrast with our own, which again only includes one modestly-sized cluster. If galaxies near Virgo-like clusters are different from those in random volumes, this could result in a sample of model galaxies with different properties. We determine the effect that this has on our analysis by comparing our predictions derived from model galaxies with those extracted from the volumes around three Virgo analogs from \citet{zakharova2024}. The \citet{zakharova2024} mock catalog uses an updated version of the GAEA models from \citet{delucia2024} that includes a different AGN feedback recipe. The models also define group membership and halo membership differently than in this paper. Despite these dissimilarities, the size ratios for filament and group galaxies change by $<30\%$ and there is very little difference for the field and cluster environments. We therefore conclude that our model predictions are relatively robust against variations in the exact selection of model galaxies and their environments. In a future work we will measure the size ratio in a much larger and more representative volume and will compare it to models from a similar volume (see \S\ref{sec:conclusion}).


\subsection{Literature Comparison}

\subsubsection{Local Cluster Survey}
\label{sec:lcs}
Because our size ratio measurements are based on techniques introduced in the Local Cluster Survey (LCS), the survey provides a useful metric of our data's consistency with existing results. The LCS consists of 9 nearby clusters within the redshift range 0.0137 $<$ z $<$ 0.0433 with GALFIT effective radii measured from Spitzer 24$\micron$ imaging, which has comparable resolution to the WISE 12$\micron$ imaging and traces the same regions of obscured star formation in a galaxy, and SDSS r-band imaging \citep{finn2018}. For their 224 galaxy sample, the group was able to find a systematic trend of lower size ratios at all but the lowest defined stellar mass bins for galaxies within cluster cores compared with those outside. They also found a correlation between B/T and size ratio, helping motivate the constraints we apply to galaxy morphology in our own VFS sample. 

In our Figure~\ref{fig:wisesizemass}, we overlay single-component size ratios from \citet{finn2018} with the size ratios divided into environments defined in Figure~\ref{fig:infallcoremap}. We calculate uncertainties according to the 68\% confidence intervals on the median points from bootstrapping. In order to correct for any potential systematic offset between the R$_{3.4}$ and R$_r$ sizes, we apply a scale factor to the LCS data according to scaling relation between the fraction of our GALFIT sizes, R$_{r}/R_{3.4}$, stellar mass.

As with the LCS, we also find that the cluster has systematically lower size ratios but with less significance than in \citet{finn2018}, especially at the lowest mass bin. There are multiple possible explanations for this finding. First and foremost, the signal in the LCS sample was heavily influenced by the Coma cluster and some of the other clusters had higher velocity dispersions than Virgo. It might therefore be the case that we would see a smaller signal in our data, as Virgo may be affecting galaxies less drastically than Coma. Second, we limit our sample to T-type greater than 0, whereas \citet{finn2018} use a $B/T<0.3$ or $n<2$ cut. Our cut includes early type spirals that would have been excluded in the LCS sample. \citet{finn2018} showed that the size ratio depends on $B/T$ and the likely different $B/T$ distributions between the two samples complicates the comparison. We also measure the size ratio between dust and 3.4\micron\ emission, whereas \citet{finn2018}, as mentioned above, base their stellar sizes on the $r$-band, which is more susceptible to radially-dependent $M/L$ variations. Finally, our definitions of external and cluster galaxies are based solely on a clustercentric radial cut, and not on a phase-space cut as performed in LCS. Despite these differences in method, we consider our results to be consistent with those from LCS. We will be analyzing a much larger sample of clusters in a future work that will further quantify the size ratio distribution in clusters.

\subsubsection{SAMI Galaxy Survey}
Still another star formation tracer is narrow-band imaging of the H$\alpha$ Balmer line, representing the 3-2 hydrogen transition line with a rest wavelength of around 656.28 nm. As an example, \cite{koopmann2004} utilized the spatial extent of H$\alpha$ for spiral galaxies in the Virgo cluster in order to correlate disk truncation with lower SFR and cluster vs. field classification. The SAMI\footnote{SAMI is a multi-object integral field spectrograph at Siding Spring Observatory.} Spectroscopic Galaxy Survey has done similar work using r$_{50,H\alpha}$/r$_{50,cont}$, which they define as the extent of ongoing star formation relative to the previous star formation (i.e., the spatial extent of the stellar population) \citep{schaefer2017, schaefer2019}. \citet{schaefer2017} found that late-type galaxies at low redshift in denser environments had steeper $H\alpha$ radial gradients, lower overall SFRs, and more centrally concentrated star formation, when compared to galaxies in lower density environments. In a later study, SAMI revealed a similar trend for galaxies in rich groups, poor groups, or the field \citep{schaefer2019}. Noticeably, they also report a slight positive relationship between the size ratios and the enhancement of SFR for low-mass galaxies.


\section{Summary}
\label{sec:conclusion}

This work set out to investigate the effects of a galaxy's place in the cosmic web on its ability to form stars. Many previous observational publications in the literature have capitalized on the stark density differences between the cluster and field environments, as these result in the strongest possible signal of environmental modification of galaxies. However, this neglects the rich intermediate structures of groups and filaments in which most galaxies actually lie. To account for this, we aimed to incorporate a more nuanced definition of environment which encompasses both filaments and groups. Doing so allowed us to understand the contributions of these regions to the modification of the spatial distribution of star formation. This more nuanced approach is important as recent simulation and observational studies \citep[e.g.][]{odekon2018, darvish2015} have suggested that filamentary networks might be the first sites of processing.

We examined galaxies in the extended regions around the Virgo cluster out to $\sim~ 12$ virial radii in projection (i.e., $\sim$ 24 Mpc) from the cluster center that has been compiled as part of the Virgo Filament Survey \citep[VFS;][]{castignani2022}. Our main observables are the effective radius of the 12$\micron$ emission and that of the 3.4$\micron$ emission, both from WISE. These were measured using GALFIT and serve as probes of the obscured star formation and stellar mass, respectively. We quantified the relative size of the star-forming and stellar disks by taking the ratio of these two sizes, then performed our analysis on a sample of galaxies restricted to those with late-type morphologies and $>10$ signal-to-noise at 12$\micron$. We divided these galaxies into separate environments either depending on their distance from the Virgo cluster core or according to their environmental classification from the VFS \citep{castignani2022}. 

Comparing the median size ratios for each environment, there is evidence to a 2.5$\sigma$ level that the galaxies in the Virgo cluster show smaller star-forming disks relative to their field counterparts. We also find, to around 1.5$\sigma$ significance, a lower median size ratio for filament galaxies compared to the field median. The remaining intermediate classifications -- rich group, poor group -- did not significantly differ from the field, which may mean that a larger sample size is needed to better disentangle any trend(s) that may lie in between cluster and field, if any. 

Overall, the statistical significance of our filament galaxy measurements compared with those of field galaxies is suggestive of some form of environmental pre-processing as galaxies pass from field to filament to cluster. While we are mindful of our filament classification also including galaxies belonging to groups and indeed potentially encompassing backsplash galaxies \citep[e.g.,][]{kuchner2022}, this finding contrasts with \citet{lee2021}, who found no evidence of either RPS or gas accretion in Virgo filaments.

We look for trends in stellar mass by separating galaxies within the cluster from those in clustercentric radii between 1.5 and 5 times the virial radius. We find no evidence of any trends existing between the median size ratios and stellar mass for either sample. However, when evaluating for such trends in our environments defined in Figure \ref{fig:sizerat_pred}, we see hints that positive correlations exist between size ratio and stellar mass for rich group galaxies. Because the SNR of the filament galaxy mass bins is so low, however, we require further data in order to fully investigate and interpret this result. Comparing between the cluster and external galaxies, we also find that the size ratio distributions between the two environments are statistically indistinguishable, despite there being hints of lower median cluster sizes emerging.

Finally, we compared our size ratio distribution to that of two SAMs \citep{xie2020} that implement tidal stripping and RPS of the hot gas, and either with or without RPS of cold gas. We apply selection criteria on the models that are as similar as possible to those used with the data. We find that our data favor the model incorporating only strangulation for most of our stellar mass bins. As both of the SAM implementations fit the quenched fraction of galaxies equally well \citep{xie2020}, the measurements of the resolved star formation disks of galaxies hold the potential to provide new constraints of the underlying model processes. 

This analysis involves galaxies in and surrounding a single nearby cluster. Recognizing this, we are reluctant to extend our data interpretation to the larger context of galaxy formation models. Our future work will build on the techniques described here to perform analogous measurements of a much larger sample drawn from a larger volume (Conger et al. in prep). Simultaneously, our team is working to characterize the filamentary environment in models in a way analogous to that in observations \citep{zakharova2023}, allowing us to probe a wider range of filamentary environments surrounding our increased sample of clusters.

 \section{Acknowledgements}
 The authors thank the International Space Sciences Institute (ISSI) in Bern, Switzerland who hosted collaboration meetings as part of the ISSI COSWEB team. They also thank the Institute for Fundamental Physics of the Universe (IFPU) in Trieste, Italy for hosting a group workshop. GHR acknowledges the support of NASA ADAP grant 80NSSC21K0641, and NSF AAG grants AST-1716690 and AST-2308127. GHR also acknowledges the hospitality of Hamburg Observatory, who hosted him during parts of this work. 
R.A.F. gratefully acknowledges support from NSF grants AST-1716657 and AST-2308127 and from a NASA ADAP grant 80NSSC21K0640.
GHR, RF, and BV thank Padova Observatory for hosting them during a team meeting. KC would like to acknowledge the support of the Maynard Redeker scholarship at the University of Kansas, as well as the travel grant awarded by the University of Kansas' Department of Physics and Astronomy. GC acknowledges the support from the Next Generation EU funds within the National Recovery and Resilience Plan (PNRR), Mission 4 - Education and Research, Component 2 - From Research to Business (M4C2), Investment Line 3.1 - Strengthening and creation of Research Infrastructures, Project IR0000012 – “CTA+ - Cherenkov Telescope Array Plus." DZ and BV acknowledge support from the INAF Mini Grant 2022 “Tracing filaments through cosmic time” (PI Vulcani).

The Legacy Surveys consist of three individual and complementary projects: the Dark Energy Camera Legacy Survey (DECaLS; Proposal ID \#2014B-0404; PIs: David Schlegel and Arjun Dey), the Beijing-Arizona Sky Survey (BASS; NOAO Prop. ID \#2015A-0801; PIs: Zhou Xu and Xiaohui Fan), and the Mayall z-band Legacy Survey (MzLS; Prop. ID \#2016A-0453; PI: Arjun Dey). DECaLS, BASS and MzLS together include data obtained, respectively, at the Blanco telescope, Cerro Tololo Inter-American Observatory, NSF’s NOIRLab; the Bok telescope, Steward Observatory, University of Arizona; and the Mayall telescope, Kitt Peak National Observatory, NOIRLab. Pipeline processing and analyses of the data were supported by NOIRLab and the Lawrence Berkeley National Laboratory (LBNL). The Legacy Surveys project is honored to be permitted to conduct astronomical research on Iolkam Du’ag (Kitt Peak), a mountain with particular significance to the Tohono O’odham Nation.

NOIRLab is operated by the Association of Universities for Research in Astronomy (AURA) under a cooperative agreement with the National Science Foundation. LBNL is managed by the Regents of the University of California under contract to the U.S. Department of Energy.
This project used data obtained with the Dark Energy Camera (DECam), which was constructed by the Dark Energy Survey (DES) collaboration. Funding for the DES Projects has been provided by the U.S. Department of Energy, the U.S. National Science Foundation, the Ministry of Science and Education of Spain, the Science and Technology Facilities Council of the United Kingdom, the Higher Education Funding Council for England, the National Center for Supercomputing Applications at the University of Illinois at Urbana-Champaign, the Kavli Institute of Cosmological Physics at the University of Chicago, Center for Cosmology and Astro-Particle Physics at the Ohio State University, the Mitchell Institute for Fundamental Physics and Astronomy at Texas A\&M University, Financiadora de Estudos e Projetos, Fundacao Carlos Chagas Filho de Amparo, Financiadora de Estudos e Projetos, Fundacao Carlos Chagas Filho de Amparo a Pesquisa do Estado do Rio de Janeiro, Conselho Nacional de Desenvolvimento Cientifico e Tecnologico and the Ministerio da Ciencia, Tecnologia e Inovacao, the Deutsche Forschungsgemeinschaft and the Collaborating Institutions in the Dark Energy Survey. The Collaborating Institutions are Argonne National Laboratory, the University of California at Santa Cruz, the University of Cambridge, Centro de Investigaciones Energeticas, Medioambientales y Tecnologicas-Madrid, the University of Chicago, University College London, the DES-Brazil Consortium, the University of Edinburgh, the Eidgenossische Technische Hochschule (ETH) Zurich, Fermi National Accelerator Laboratory, the University of Illinois at Urbana-Champaign, the Institut de Ciencies de l’Espai (IEEC/CSIC), the Institut de Fisica d’Altes Energies, Lawrence Berkeley National Laboratory, the Ludwig Maximilians Universitat Munchen and the associated Excellence Cluster Universe, the University of Michigan, NSF’s NOIRLab, the University of Nottingham, the Ohio State University, the University of Pennsylvania, the University of Portsmouth, SLAC National Accelerator Laboratory, Stanford University, the University of Sussex, and Texas A\&M University.

BASS is a key project of the Telescope Access Program (TAP), which has been funded by the National Astronomical Observatories of China, the Chinese Academy of Sciences (the Strategic Priority Research Program “The Emergence of Cosmological Structures” Grant \# XDB09000000), and the Special Fund for Astronomy from the Ministry of Finance. The BASS is also supported by the External Cooperation Program of Chinese Academy of Sciences (Grant \# 114A11KYSB20160057), and Chinese National Natural Science Foundation (Grant \# 12120101003, \# 11433005).
The Legacy Survey team makes use of data products from the Near-Earth Object Wide-field Infrared Survey Explorer (NEOWISE), which is a project of the Jet Propulsion Laboratory/California Institute of Technology. NEOWISE is funded by the National Aeronautics and Space Administration.

The Legacy Surveys imaging of the DESI footprint is supported by the Director, Office of Science, Office of High Energy Physics of the U.S. Department of Energy under Contract No. DE-AC02-05CH1123, by the National Energy Research Scientific Computing Center, a DOE Office of Science User Facility under the same contract; and by the U.S. National Science Foundation, Division of Astronomical Sciences under Contract No. AST-0950945 to NOIRLab.




\bibliography{report_template.bbl}

\appendix

\section{unWISE Background Sensitivity}
\label{sec:Appendixbackground}

\begin{figure}[h]
\includegraphics[scale=0.295]{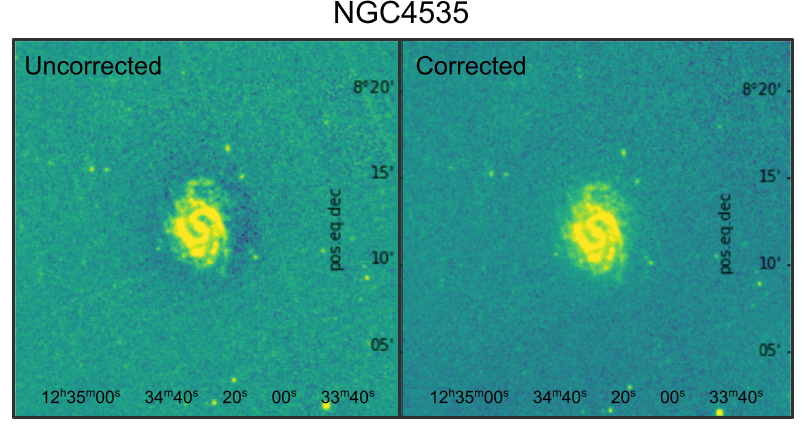}
\centering
\caption{An example massive galaxy from the unWISE 12\micron\ imaging showing an oversubtraction halo (left). The right postage stamp is of that same galaxy with corrected image processing.}
\label{fig:oversubtraction}
\end{figure}

As described in \S\ref{sec:irimaging}, cutout images of the most massive galaxies of the sample displayed strange divots surrounding the galaxy's perimeter (see Figure \ref{fig:oversubtraction}, left). We later found these were the result of background oversubtraction and corrected (D. Lang, priv comm) the masking routine which mistakenly identified the outskirts of these galaxies as background. This correction eliminated the divots, as in Figure \ref{fig:oversubtraction}, right.

\section{GALFIT Background Sensitivity}
Many of our galaxy postage stamps include non-central objects that will otherwise contaminate the background flux unless masked. As the fraction of pixels can become large, we determined the maximum masked-to-total pixel ratio (MTR) before GALFIT parameters start to be affected.

\begin{figure}[h]
\includegraphics[scale=0.48]{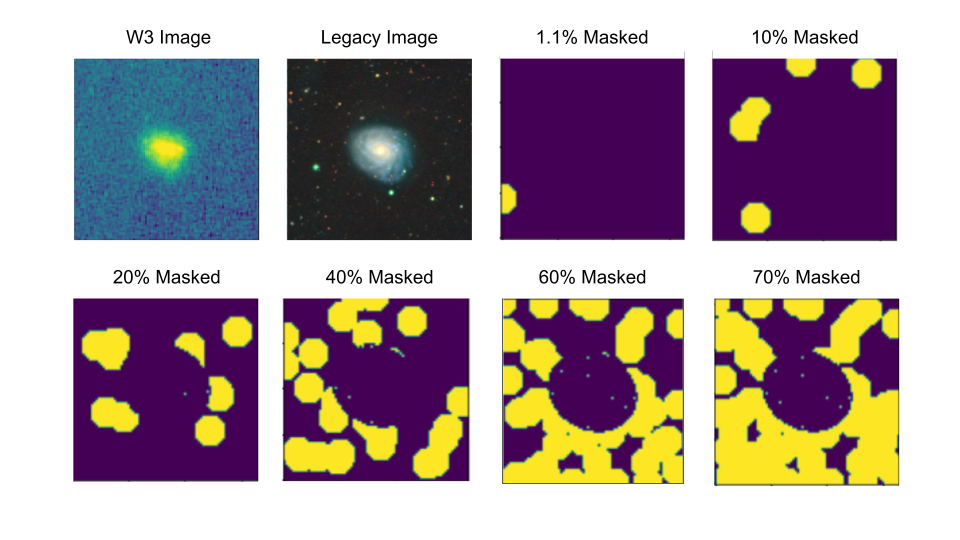}
\centering
\caption{Mask examples for the cases with the MTR = 0.01, 0.10, 0.20, 0.40, 0.60, and 0.70. The upper left panels showcase the 12$\micron$ and Legacy Survey optical images for the test galaxy, NGC3364. Note that the region enclosing the galaxy is largely unmasked aside from individual pixels with maskbit values not exclusively marked as belonging to the SGA galaxy. See https://www.legacysurvey.org/dr9/bitmasks/ for further details.}
\label{fig:maskex}
\end{figure}

Testing this idea first involves identifying a galaxy cutout with a relatively smooth background, which we select to be NGC3364 in the 12$\micron$ band. We then run GALFIT to acquire baseline parameters and subsequently insert circular masked regions of fixed radius and randomly-generated center coordinates. We create a series of mask images within increasing MTR until a maximum value of 0.70, with each consecutive image containing one additional circular aperture. Example panels of what these masks look like are in Figure \ref{fig:maskex}. Each MTR is run with 10 iterations, varying the location of the masked pixels. We tested for cases both with and without convolution, as well as with and without restricting mask apertures to the region outside of the galaxy bounds as defined by the Siena Galaxy Atlas\footnote{https://www.legacysurvey.org/dr9/bitmasks/} masking routine. We only report on the combination of convolution and no galaxy masking below, as this pairing best reflects the actual masking procedures we apply in this paper.

\begin{figure}[h]
\includegraphics[scale=0.32]{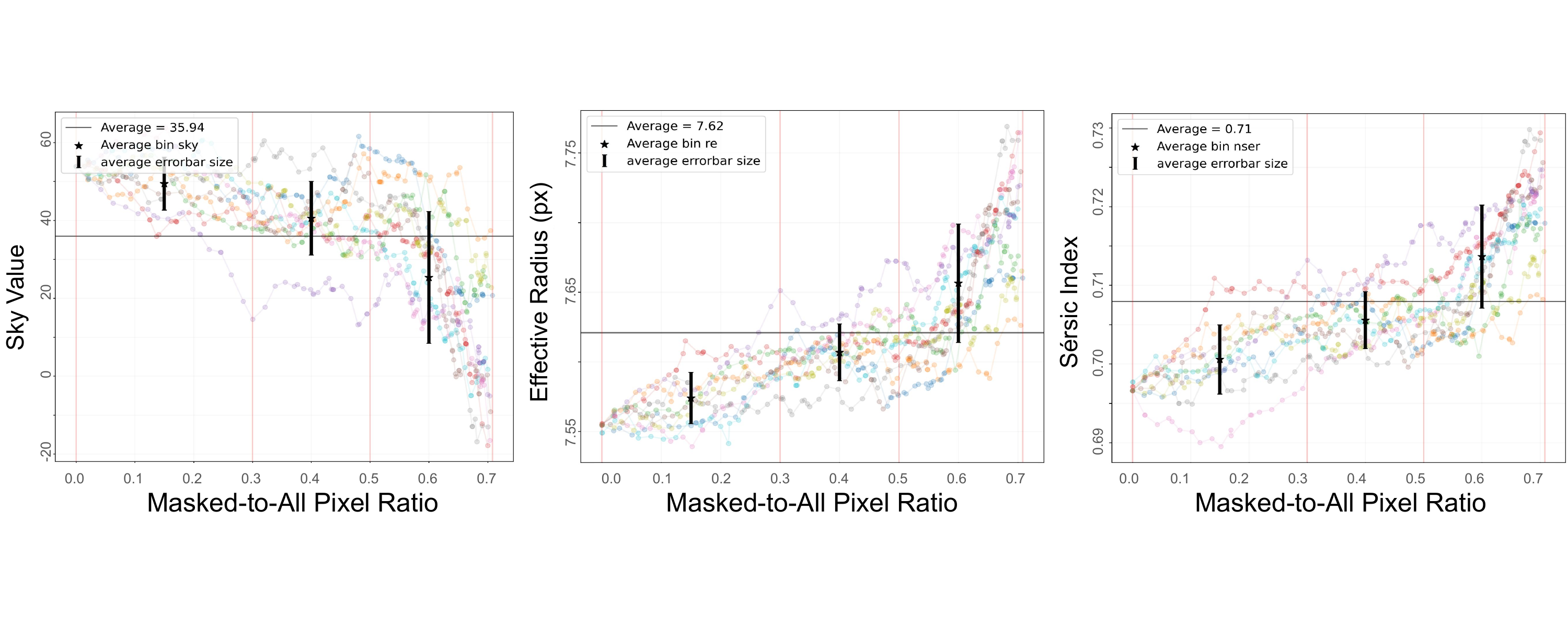}
\centering
\caption{GALFIT output parameters for NGC3364 run with convolution, with each color representing a single iteration of the test described above. We plot the variations for three parameters: the sky value (left), the effective radius (middle) and the Sérsic index (right). The vertical red lines represent each MTR bin: 0$<$MTR$<$0.3, 0.3$<$MTR$<$0.5, and 0.5$<$MTR$<$0.7. The black markers denote the mean parameter value within each of these bins, with error bars being the average uncertainty that GALFIT assigned to each value. The horizontal gray lines trace the average parameter value across all bins.}
\label{fig:galfitbackground}
\end{figure}

The results of this analysis are in Figure \ref{fig:galfitbackground}, displaying the sky value, effective radius, and Sérsic index variations as the percentage of background pixels increase. Both the scatter and deviation from the no-mask case increases with increasing MTR, with positive correlations for effective radius and Sérsic index and a negative correlation for the sky value. The error bars, which represent the average GALFIT parameter uncertainty for all data points within the red vertical bar bins, also convey this narrative. For the effective radius and Sérsic index cases, the uncertainty widths increases by nearly a factor of 10 between columns, while the sky uncertainties increase by a factor of 2. GALFIT is therefore, understandably, sensitive to a reduction in available background pixels with which to estimate the sky. 

We conclude that there are only small statistical deviations from the baseline for MTR$<0.4-0.5$. For MTRs lower than 0.5, there are also only marginal differences between bins for both the Sérsic index and effective radius parameters. Indeed, the Sérsic index remains within the same morphological classification, while the effective radius changes on the order of only a fraction of a pixel. We also see that, as evidenced by the plots in Figure \ref{fig:galfitbackground}, sky value estimates do not dramatically affect the output parameters relevant to our galaxy evolution analysis. Ultimately, we use these plots to inform our decision to restrict our sample MTRs to 0.5 in order to preserve parameter robustness while simultaneously allowing flexibility for more aggressive masking of galaxy fields with multiple background sources. 

\end{document}